\documentstyle[a4]{article}

\title{\bf On the Canonical Structure of the De Donder-Weyl Covariant
Hamiltonian Formulation of Field Theory  \\ I. Graded Poisson brackets
and equations of motion}
\author{Igor V. Kanatchikov \thanks{e-mail:
igor@thphys.physik.rwth-aachen.de}\\ Institut f\"{u}r Theoretische
Physik \\ RWTH Aachen \\ 52056 Aachen, F.R. Germany}
\date{}

\begin{document}

\maketitle

\vspace*{-3.4in}

\begin{flushright}
PITHA 93/41 \\ November 1993\\
\end{flushright}

\vspace*{3.5in}

\begin{abstract}
As opposed to the conventional field-theoretical Hamiltonian formalism,
which requires the \linebreak[3] space+time decomposition and leads to
the picture of a field as a mechanical system with infinitely many
degrees of freedom, the De Donder-Weyl (DW) Hamiltonian canonical
formulation of field theory (which is known for about 60 years) keeps
the space-time symmetry explicit, works in the {\em finite} dimensional
analogue of the phase space and leads to  the Hamiltonian and
Hamilton-Jacobi formulations of field equations  in terms of {\em
partial} derivative equations. No field quantization procedure based on
this "finite dimensional" covariant canonical formalism is known.  As a
first step in this direction we consider the appropriate generalization
of the Poisson bracket concept to the DW Hamiltonian formalism and the
expression of the DW Hamiltonian form of field equations in terms of
these generalized Poisson brackets. Starting from the
Poincar\'{e}-Cartan form of the multidimensional variational calculus
we argue that the analogue of the Poisson brackets is defined on {\em
forms} of different degrees and is related to the Schouten-Nijenhuis
bracket of the corresponding multivector fields. The forms generalize
the dynamical variables (functions) of mechanics and the multivector
fields generalize the Hamiltonian vector fields associated with
dynamical variables.  The corresponding map between forms and
multivectors is determined by the "polysymplectic" $(n+1)$-form (given
by the Poincar\'{e}-Cartan form) which we consider as the analogue of
the symplectic form in the DW Hamiltonian formalism for fields.  The
space of "Hamiltonian forms" equipped with the exterior product and our
Poisson bracket is shown to constitute the {\em Gerstenhaber} graded
algebra.  We also demonstrate that the Poisson bracket of any form with
the $n$-form $H\widetilde{vol}$, where $H$ is the DW Hamiltonian
function, generates its exterior differential and this enables us to
write the DW Hamiltonian field equations in the bracket form.  Finally,
we present few simple examples illustrating how the formalism works in
some field-theoretical models, and also briefly discuss the relation to
the conventional Hamiltonian description of fields.

\end{abstract}

\renewcommand{\baselinestretch}{1.3}
\normalsize\large

\newcommand{\beq}{\begin{equation}}
\newcommand{\eeq}{\end{equation}}
\newcommand{\rbox}[2]{\raisebox{#1}{#2}}
\newcommand{\xx}[1]{\raisebox{1pt}{$\stackrel{#1}{X}$}}
\newcommand{\xxi}[2]{\raisebox{1pt}{$\stackrel{#1}{X}$$_{#2}$}}
\newcommand{\lie}[1]{$\cal L$$_{\stackrel{#1}{X}}$}
\newcommand{\pbr}[2]{\{ #1 , #2 \}}
\newcommand{\der}{\partial}
\newcommand{\oo}{$\Omega$}
\newcommand{\inn}{\hspace*{2pt}\raisebox{-1pt}{\rule{6pt}{.3pt}\hspace*
{0pt}\rule{.3pt}{8pt}\hspace*{2pt}}}
\newcommand{\sro}{Schr\"{o}dinger\ }
\newcommand{\bm}{\boldmath}
\newcommand{\vol}{\widetilde{vol}}

\newpage

\section{Introduction}

There are known two different approaches to the Hamiltonian formulation
of field theory: the first is built on the infinite dimensional
"instantaneous" phase space and implies certain space+time
decomposition while the second is formulated on the finite dimensional
analogue of the phase space and is manifestly space-time covariant.
Both are based on certain extensions of the structures of classical
analytical mechanics and one dimensional variational calculus.  The
first  approach singles out the time dimension and treats a field as a
mechanical system with continually infinite number of degrees of
freedom. The generalized coordinates are the values of fields $y^a$ at
each point of the space at a given instant of time $y^a({\bf x})$, and
the generalized canonical momenta are defined from the Lagrangian
density $L$ to be $p_a({\bf x})=\partial L/\partial(\partial_ty^a({\bf
x}))$ as in mechanics. This is, of course, a well known conventional
treatment used for example when canonically quantizing the fields.
Recent discussion of the covariant version of this approach may be
found for example in  \cite{crnkovic}.\\

   The second approach, that we are concerned with in this paper,
originates from the approaches to the multidimensional variational
problems due to De Donder \cite{De Donder}, Carath\'{e}odory
\cite{Carath29}, Weyl \cite{Weyl35} and some others (see for example
\cite{Rund}) and \cite{Kastrup83} for a review). It is entirely
space-time covariant because a field is treated as a sort of
generalized Hamiltonian dynamical system with many "times". This means
that both space and time enter the formalism on an equal footing as
variables over which a field "evolution" proceeds. By "evolution" one
means here not merely a time evolution from the given Cauchy data, as
usual, but any space-time development or variation of a field. In this
approach\footnote{In fact, we consider here only the simplest
particular representative of a whole variety of finite dimensional
canonical theories for fields which differ by an extra "Lepagean" term
added to the canonical Hamilton-Poincare-Cartan form (see eq. (2)); an
excellent survey may be found in \cite{Kastrup83}, see also  \cite
{Gotay ext,Gotay multi1}.  This particular case is sometimes called the
De Donder-Weyl (DW) canonical theory and we also will use this term.}
the generalized coordinates are the field variables $y^a$ to which a
{\em set} of canonically conjugate momenta $p^i_a:=\partial
L/\partial(\partial_iy^a)$ is associated. The De Donder-Weyl
Hamiltonian function is defined as $H_{DW}:=p^i_a\partial_iy^a - L $
provided the corresponding generalized Legendre transform is regular.
Note that unlike the first approach, the Hamiltonian function is
scalar, but its direct physical interpretation, if there is any, is not
evident.  The generalization of the extended phase space of mechanics
in this approach is a finite dimensional phase space of the variables
$(y^a,p^i_a,x^i)$ which replaces the infinite dimensional phase space
of the instantaneous approach. As a consequence, the Euler-Lagrange
field equations may be written in the corresponding Hamiltonian form in
an entirely covariant way (see eqs. (7) below) and in terms of {\em
partial} derivative equations.  The corresponding Hamilton-Jacobi
theory (see for example
\cite{Rund,Kastrup83}) is also
formulated in terms of the covariant  partial differential equation as
opposed to the first approach leading to the functional derivative
equation.  The connection between the instantaneous and the covariant
finite dimensional formulations was studied recently in detail by Gotay
\cite{Gotay multi2} (see also the book \cite{Gotay ea}).\\

     Despite all of the attractive features of the second treatment
which look especially relevant in the context of general relativity and
string theory, there is surprisingly small number of its applications
to relativistic field theories
\cite{Kij+Tul,Rieth,Grigore sour}, gauge fields \cite{Kij84,Sardan},
classical bosonic string \cite{Kastr+Rinke,Nambu80,Beig,Kanatch,Grigore
ngoto,Wulf} and general relativity \cite{Francav,Horava} in the
literature (see also \cite{Gotay ea}). In particular, it remains
unclear till now how to develop a field quantization starting from this
finite dimensional Hamiltonian treatment on the classical level and
whether it is possible or has a sense at all.  Indeed, is it really
necessary to split at first the space-time in order to obtain the
Hamiltonian formulation, and then to quantize a field according to
standart prescriptions of quantum theory and to prove the procedure to
be consistent with the relativistic symmetries, or it is possible
instead to develop a field quantization based on the finite dimensional
covariant Hamiltonian framework and then  obtain the space-time
splitted results, as it is operationally required, from the manifestly
covariant quantum field theory?  Another related question is whether
there exists a quasiclassical transition from some formulation of a
quantum field theory to the Hamilton-Jacobi equations corresponding to
the finite dimensional canonical formulations of classical fields. \\

   The problem of field quantization based on the finite dimensional
Hamiltonian formalism, which is the main motivation of our study, was
shortly discussed in thirties by Born \cite{Born34} and Weyl
\cite{Weyl34}.  In early seventies a considerable progress was made in
understanding the differential geometric structures of the De
Donder-Weyl canonical formalism \cite{Garcia,Gold+Stern,Kij ea,Gaw}
(see also Dedecker \cite{Dedecker}, who studied more general canonical
theories and the recent paper by Gotay \cite{Gotay ext} for a
subsequent development), however the attempts
\cite{Herm lie,Garcia,Kij ea,Gaw}
to approach from this viewpoint a quantum field theory did not lead to
any new formulation but have established some links with the
conventional one which is based on the instantaneous Hamiltonian
formalism.  More recently the attempt to construct a quantum field
theoretical formalism based entirely on the finite dimensional DW
canonical theory was reported by G\"{u}nther in \cite{Guenther87b} who
used his own \cite{Guenther87a} geometrical version of the DW canonical
theory, the "polysymplectic Hamiltonian formalism".  Unfortunately, the
ideas of his brief report \cite{Guenther87b} were not developed to the
extend which would allow us to compare the outcome with something known
from the conventional quantum field theory.  \\

	The main obstacle in the direction of a "finite dimensional
field quantization" seems to be the lack of an appropriate
generalization or analogue of the Poisson brackets in the classical
canonical theories under discussion. Within the DW Hamiltonian theory,
the brackets of the $(n-1)$-forms corresponding to observables in field
theory were proposed in \cite {Gold+Stern,Herm lie, Kij ea,Gaw}, but
the related construction proved to be too restrictive to reproduce the
algebra of observables in the theories of sufficiently general type and
were not appropriate for representing the canonical Hamiltonian field
equations in the bracket form. Another approaches  due to Good
\cite{Good}, Edelen \cite{Edelen} and G\"{u}nther \cite{Guenther87a}
enable one to write the canonical equations in the bracket form,
however, the group theoretical properties of their brackets are not
evident.\\

   The purpose of the present study is to develop those elements of the
finite dimensional canonical formalism for fields which are essential
for the canonical quantization. Some of the questions (chosen more or
less randomly) which arise as soon as  we are trying to quantize a
field theory (let's say in the Schr\"{o}dinger picture) proceeding from
a finite dimensional canonical framework are the following: (i) which
variables may be considered as the canonically conjugate ones and in
which sense (notice, that the number of generalized coordinates $y^a$
and generalized momenta $p^i_a$ is different!), (ii) what are the
Poisson brackets corresponding to the finite dimensional canonical
formalism, (iii) if these brackets exist, do they generate the
equations of motion, like in mechanics, if one of the arguments is the
DW Hamiltonian function, and lastly (iv) which operation replaces in
the bracket representation of the DW canonical equations the total time
derivative and describes "evolution" in the sense explained above?  \\

  To approach these and other related questions, we suggest to proceed
from the most fundamental object of any canonical theory, the
Hamilton-Poincar\'{e}- Cartan (HPC) $n$-form ($n=$space-time
dimension), and try to develop the subsequent formalism by searching
for the proper generalizations of the corresponding  elements of the
canonical formalism of mechanics (see e.g.
\cite{Abr+Marsden, Arnold}) to the finite-dimensional canonical formulation
of field theory. \\

   The structure of the paper is the following. At first we recall in
Sect. 2 how the DW Hamiltonian  field equations comes directly from the
canonical HPC form. This consideration indicates
that in field theory the suitable generalization of the notion
of the canonical Hamiltonian vector field is the multivector field
of degree $n$, and also suggests the analogue of the symplectic form to
be certain $(n+1)$-form (see eq. (10)), called {\em polysymplectic},
which is obtained from the HPC form. Then, in Sect. 3 we suggest the
generalization of the principle of preservation of the symplectic
structure, which is a cornerstone of the classical canonical
transformation theory, to field theory. This generalization involves
the extension of the notion of Lie derivative giving sence to the Lie
derivative with respect to a {\em multi}vector field. The symmetry
postulate that the generalized Lie derivative of the polysymplectic
form with respect to the Hamiltonian multivector field vanishes, turns
out to be consistent with the canonical DW equations. Moreover, using
the generalized Lie derivative allows us to define the bracket
operation of both the generalized Hamiltonian fields which are {\em
vertical} (in the sense defined in Sect. 3) multivectors of various
degrees $p=1,...,n$  and the generalised Hamiltonian functions which
are ({\em horizontal}) forms of degrees $q=0,...,n-1$. The first
bracket turns out to be the Schouten-Nijenhuis (SN) bracket of
multivectors and the second one, which is the analogue of the Poisson
bracket acting on Hamiltonian forms, is related to the SN bracket and
to the polysymplectic form in just the same way as the Lie bracket of
Hamiltonian vector fields is related to the Poisson bracket of
Hamiltonian functions in mechanics. We also show that the space of
Hamiltonian forms equipped with the exterior product and the Poisson
bracket of forms becomes essentially the {\em Gerstenhaber algebra}
\cite{Gerstenhaber}. The generalized
Poisson bracket of Hamiltonian forms (see Sect. 3 for an explanation of
this term) is used in Sect. 4 for  representing the DW Hamiltonian
field equations of $(n-1)$-forms in the bracket form. As a by-product
we also discuss the proper generalization of the notions of  integral
of motion and canonically conjugate variables to the DW canonical
formalism. Then, in Sect.5  we enlarge the set of Hamiltonian forms by
adding the $n$-forms. It enables us  to express the
equations of motion of Hamiltonian forms of any degree in the bracket
form. However, this enlargement implies also an extension of the space
of Hamiltonian multivector fields and the algebraic closure of this
enlarged space is argued to involve both the (vertical) vector-valued
one-forms corresponding to $n$-forms and the multivector-valued forms
of higher degrees. Thus the algebraic closure of the space of
generalized Hamiltonian "vector" fields leads to the problem of
embedding of both the Schouten-Nijenhuis and the
Fr\"{o}licher-Nijenhuis graded Lie algebras in some larger algebraic
structure which includes the multivector-valued forms of all possible
degrees. Solving this problem remains beyond the scope of our paper.
Finally, some simple applications of our Poisson brackets to
interacting scalar fields, electrodynamics and the Nambu-Goto string
are presented in Sect. 6, and a general discussion including some
remarks on the connections with the conventional Hamiltonian
formulation may be found in the concluding Sect. 7.   \\


\section{Poincar\'{e}-Cartan form and the De Donder-Weyl Hamiltonian field
equations}

Given a first order multidimentional variational problem
\begin{equation}
\delta \int L(y^a, \partial_iy^a, x^i)\widetilde{vol}=0,
\end{equation}
where $(y^a), 1\leq a\leq m$ are field variables,
$(x^i), 1\leq i\leq n$ are space-time variables, and $\widetilde{vol}:=
dx^i\wedge...\wedge dx^n $ \footnote{In order to simplify formulae we
imply in the following that the coordinates on the x-space are choosen
such that the metric determinant $|g|=1$.}, it is known that the
Hamilton-Poincar\'{e}-Cartan (HPC) fundamental $n$-form is defined
within the De-Donder-Weyl (DW) approach to multidimensional variational
problems as (see for example \cite{Kastrup83, Gotay ea} and
\cite{Gold+Stern})

\[\Theta_{DW}=p^i_a\wedge dy^a\wedge\partial_i\inn\vol - H_{DW}\vol  \]

so that its exterior differential is given by
\begin{equation}
\Omega_{DW}=dp^i_a\wedge dy^a\wedge\partial_i\inn\widetilde{vol}-dH_{DW}
\wedge\widetilde{vol}. \\
\end{equation}
Here $p^i_a:=\partial L/\partial(\partial_iy^a)$ are the DW  canonical
momenta  and
\begin{equation}
H_{DW}(y^a,p^i_a, x^i):=p^i_a\partial_iy^a-L
\end{equation}
is the DW Hamiltonian function. The symbol \inn\  denotes the interior
product of a (multi)vector on the left and a form on the right.  In the
following we will omit the subscript DW, but the quantity $H$ which we
call the (DW) Hamiltonian function should not be confused with usual
Hamiltonian which is related to energy.\\

The form $\Omega_{DW}$ contains in a sense all the information about
field dynamics. In particular, one can derive the appropriate
Hamiltonian form of field equations directly from $\Omega_{DW}$.
Indeed, the solutions of the variational problem (1) may be considered
as n-dimensional distributions in the extended DW phase space with the
coordinates
\[Z^M:=(y^a, p^j_b, x^i).\]
These distributions one can describe by the n-multivector (or n-vector,
in short) field \xx{n}:
\begin{equation}
\mbox{$\stackrel{n}{X}:=\stackrel{n}{X}$$^{M_1...M_n}(Z)\,\partial_{M_1}
\wedge...\wedge\partial_{M_n}$ }
\end{equation}
representing their tangent n-planes. Then the condition on
$\stackrel{n}{X}$ to give the classical extremals is that the form
$\Omega$$_{DW}$  should vanish on \xx{n} (cf. e.g.
\cite{Kastrup83,Gotay multi1,Gotay ea,Kij+Tul,Gold+Stern,Herm lie}), i.e.

\begin{equation}
\mbox{\xx{n}\inn\ $\Omega_{DW}=0.$ } \\
\end{equation}
The n-vector field $\stackrel{n}{X}$ naturally generalizes the velocity
field of the canonical Hamiltonian flow in classical mechanics
corresponding to $n=1$ to field theory, which corresponds to $n > 1$.
Eq. (5) gives the components of the n-vector annihilating the
$(n+1)$-form $\Omega_{DW}$ and together with the following natural
parametrization of the components of $\stackrel{n}{X}$:
\begin{equation}
\mbox{$\stackrel{n}{X}$$^{M_1...M_n}$}=\frac{\partial(Z^{M_1},...,Z^{M_n})}
{\partial(x^1,...,x^n)},
\end{equation}
leads to the set of equations\footnote{In fact, the components
\xx{n}$^{aii_1 ... i_{n-2}}_{\cdot a}$ of the n-vector \xx{n} yield also the
third equation which may be shown to be a consequence of eqs. (7,8).
Thus the information about the classical dynamics of field is
essentially encoded in the "vertical", as we call them below,
components $X$$^{ai_1 ... i_{n-1}}$ and $X$$^{ii_1...i_{n-1}}_{a}$.
This is the observation which motivates our  construction in Sect.3.}

\begin{eqnarray}
\partial_ip^i_a&=-\partial_aH&:=-\frac{\partial H}{\partial y^a}, \\
               &             &          \nonumber \\
\partial_iy^a&= \partial^a_iH&:=\frac{\partial H}{\partial p^i_a},
\end{eqnarray}
which we will refer to as the (DW) Hamiltonian field equations. They
are equivalent to the Euler-Lagrange equations one gets from the
variational problem (1) and are the simplest field theoretic
generalization of the canonical Hamilton's equations of motion.


\newcommand{\ff}[1]{\raisebox{1pt}{$\stackrel{#1}{F}$}}
\newcommand{\nbr}[2]{{\bf[}#1 , #2{\bf ]}}
\newcommand{\dvol}[1]{\partial_{#1}\inn \vol }

\section{Polysymplectic form, Hamiltonian multivector fields and forms and the
generalization of the Poisson bracket.}

   Our task in this section is to find the appropriate generalization
of the basic structures of classical Hamiltonian mechanics, such as the
Hamiltonian vector fields and functions, the symplectic structure,
Poisson brackets etc., to the DW Hamiltonian formulation of field
theory.\\

   Notice first that, as it follows from (5), the Hamiltonian field
equations (7),(8) can be derived also from the condition
\begin{equation}
\mbox{\rbox{1pt}{$\stackrel{n}{X}$$^{v}$}\inn\ $\Omega^{v} = d^{v}H,$}
\end{equation}
where the superscript $^v$ shows that we take  a vertical'  part of the
quantity. We call {\em vertical} the variables $z^V=(y^a,p^i_a)$ and
the corresponding subspace of the extended DW phase space and {\em
horizontal} the space-time variables $(x^i)$; the index $V$ corresponds
to vertical variables. Further, the $p$-multivector is called vertical
if it has one vertical and $(p-1)$ horizontal indices,
i.e.

\[\mbox{\xx{p}$^{v}:=$ \xx{p}$^{Vi_1...i_{p-1}}(Z)
\der_{V}\wedge\der_{i_1}\wedge...\wedge\der_{i_{p-1}},$}\]
and the vertical exterior differential $d^v$ of any form $\omega$ is
defined as
\[d^v \omega := dz^V \wedge \der_V \omega,      \]
so that, in particular,

\[d^{v}H = \der_aH dy^a + \der^a_iH dp^i_a.  \]
In the following we will also use the notion of {\em horizontal}
$p$-forms which are defined to have a form
\[\ff{p}:=F_{i_1 ... i_p}(Z) dx^{i_1}\wedge ...\wedge dx^{i_p}.       \]
Finally, the form $\Omega^v$ in eq.(9) is defined as
\beq
\Omega^{v} := dp^i_a\wedge dy^{a} \wedge\der_{i} \inn\ \widetilde{vol}. \\
\eeq
so that it is given by the vertical exterior differential of the
vertical part of the HPC form:
\[\Omega^v := d^v \Theta^v,\]

\[\Theta^v := p^i_a dy^a \wedge \dvol{i}.\]
In the following, the closed $(n+1)$-form $\Omega ^v$ will be denoted
as $\Omega$,  and we shall call it the {\em polysymplectic} form
adopting the term introduced in a similar context earlier
\cite{Guenther87a}. We will also omit all the superscripts $^v$ of the
multivectors, since all of them appearing in the following will be
taken to be vertical, unless the opposite will explicitly be stated.
Note also that the polysymplectic form is related to the HPC form in
exactly the same way as the symplectic form in mechanics is related to
the HPC form of the $1$-dimensional variational problem. \\

Let us recall now (see for details \cite{Abr+Marsden,Arnold}) that the
structures of classical Hamiltonian mechanics are contained essentially
in a single statement that Lie derivative of a symplectic form $\omega$
with respect to the vertical vector fields $X$ generating the
infinitesimal canonical transformations vanishes: $\pounds$$_X \omega
=0$.  Since $\omega$ is closed, it implies locally that $X_f \inn\
\omega =df$ for some function $f$ of the phase space variables. If this
equality holds globally, the vector field $X_f$ is said to be
(globally) Hamiltonian vector field associated with the Hamiltonian
function $f$. When $f$ is taken to be the canonical Hamilton's function
$H$, the equations of the integral curves of $X_H$ (the canonical
Hamiltonian vector field) reproduce Hamilton's canonical equations of
motion. \\

	From the previous considerations one could already notice that
in field theory the $n$-vector field \xx{n} associated with the DW
Hamilton's function as in eq.(9) is similar to the canonical
Hamiltonian vector field in mechanics and the $(n+1)$-form $\Omega$ is
similar to the symplectic 2-form.  To pursue this parallel further, let
us introduce the generalized Lie derivative -- \lie{p} -- with respect
to a multivector field of degree $p$ and postulate, as a fundamental
symmetry principle, that

\beq
\mbox{\lie{n} $\Omega=0.$}
\eeq
We define the generalized Lie derivative of any form $\omega$ with
respect to the multivector field \rbox{1pt}{$\stackrel{p}{X}$} of
degree $p$ (not necessarily vertical) by the formula

\beq
\mbox{$\cal L$\rbox{1pt}{$_{\stackrel{p}{X}}$} $\omega :=$
\rbox{1pt}{$\stackrel{p}{X}$}\inn\
$d\omega-(-1)^p \, d$(\rbox{1pt}{$\stackrel{p}{X}$}\inn\ $\omega)$}
\eeq
which  is the simplest generalization of the Cartan formula relating
the Lie derivative of a form along the vector field to the exterior
derivative and the inner product with the vector; this relation is
recovered when $p=1$.  Note however that, unlike the $p=1$ case, the
operation $\cal L$$_{\stackrel{p}{X}}$ does not preserve the degree of
a form it acts on: it maps q-forms to $(q+1-p)$-forms. If \xx{p} is
vertical, then the definition of \lie{p} is modified by replacing the
exterior differentials in eq.(12) by the vertical ($d^v$) ones.\\

   Since $\Omega$ is closed with respect to the vertical exterior
differential, from the symmetry postulate, eq.(11), and the definition
of the generalized Lie derivative, eq.(12), it follows
\beq
\mbox{$d^{v}$ (\rbox{1pt}{$\stackrel{n}{X}$} \inn\ $\Omega)=0,$}
\eeq
{\rm so that locally one can write}

\begin{equation}
\mbox{\rbox{1pt}{$\stackrel{n}{X}$}\inn\ $\Omega=d^{v}$\rbox{1pt}
{$\stackrel{0}{F}$} }
\end{equation}
for some $0$-form $\stackrel{0}{F}$  depending on the phase space
variables $Z^M$. By analogy with mechanics, if such a form exists
globally, we call \xx{n} the (globally) Hamiltonian n-vector field
(associated with the Hamiltonian $0$-form \ff{0}) while the \xx{n}
satisfying eq. (13) is called locally Hamiltonian. We see
from eqs. (9) and (13) that our postulate, eq. (11), together with the
definition of the generalized Lie derivative in eq. (12) leads to the
correct DW Hamiltonian field equations if the the $0$-form \ff{0} at
the r.h.s. of eq. (13) is taken to be the DW Hamiltonian $H$.\\

\newcommand{\nbrpq}[2]{\nbr{\xxi{#1}{1}}{\xxi{#2}{2}}}
\newcommand{\lieni}[2]{$\cal L$$_{\stackrel{#1}{X}_{#2}}$  }

Given two locally Hamiltonian n-vector fields it is natural to  define
their bracket as

\beq
\mbox{\nbrpq{n}{n} \inn\ \oo\, := \lieni{n}{1}(\xxi{n}{2}\inn\ \oo) , }
\eeq
which is obviously in accordance with the invariance property we have
postulated in eq. (11). From the definition in eq. (15) it follows

\beq
\mbox{$d^{v}$(\nbrpq{n}{n} \inn\ \oo) = 0,}
\eeq
so that \nbrpq{n}{n}\ is also locally Hamiltonian.  However, this
bracket does not map n-vector fields to n-vector ones; instead it mixes
the multivectors of different degrees. Moreover, as the counting of
degrees in eq. (15) gives $deg$(\nbrpq{n}{n}) = $2n-1$, the bracket
vanishes identically if $2n-1>n$, i.e. for $n>1$. These observations
indicate that the multivectors and, therefore, the forms of various
degrees should come into play.\\

Thus, given the polysymplectic $(n+1)$-form \oo, we shall define the
set of {\em locally Hamiltonian} ($LH$) multivector fields as the set
of {\em vertical} $p$-vector fields \xx{p}, $1\leq p \leq n$,   for
which
\beq
\mbox{\lie{p} \oo\ $ = 0.$}
\eeq
Then the p-vector fields are defined to be {\em Hamiltonian} if there
exist {\em horizontal} q-forms
\ff{q}:=$F_{i_1...i_q}(Z) dx^{i_1}\wedge ...
\wedge dx^{i_q}$, $0 \leq q \leq (n-1),$ such that

\beq
\mbox{\xxi{p}{F}\inn\ \oo \, = $ d^v$\ff{q},}
\eeq
where $p=n-q$. In the following we call the forms $\ff{q}$ the {\em
Hamiltonian forms} and the multivector fields $\xx{p}$$_{F}$ the {\em
Hamiltonian multivector fields} generated by (or associated with) the
forms $\ff{q}$.  The set of Hamiltonian forms extends to field theory
the notion of Hamiltonian functions or dynamical variables in
mechanics. The inclusion of forms of various degrees is motivated by
the fact that the dynamical variables of interest in field theory in
$n$ dimensions can be the forms of any degree  $p \leq n$ (the
$n$-forms are incorporated in Sect.5). It should be noted that, in
contrast with mechanics, eq. (18) imposes rather strong restriction on
the functional dependence of the components of Hamiltonian forms on the
DW momenta (see eq.  (29) below for the case of $(n-1)$-forms). \\

	The bracket of two locally Hamiltonian fields may be defined
now similarly to eq. (15):
\beq
\mbox{\nbrpq{p}{q}\inn\ \oo\ := \lieni{p}{1}(\xxi{q}{2}\inn\ \oo), }
\eeq
and it is easy to show that it maps the {\em LH} fields to {\em LH}
ones. This bracket (i) generalizes the Lie bracket of vector fields,
(ii) its degree is easily found to be
\beq
\mbox{$deg($\nbrpq{p}{q}$) = p+q-1$, }
\eeq
(iii) it can be both odd and even
\[\mbox{\nbrpq{p}{q} = $-(-1)^{(p-1)(q-1)}$ \nbr{$\xxi{q}{2}$}{$\xxi{p}{1}$}}\]
and, finally, (iv) it fulfils the graded Jacobi identities
\begin{eqnarray}
\mbox{$(-1)^{g_1 g_3}$\nbr{\xx{p}}{\nbr{\xx{q}}{\xx{r}}}} &+&
                                           \mbox{\hspace*{15em}} \nonumber \\
\mbox{$(-1)^{g_1 g_2}$\nbr{\xx{q}}{\nbr{\xx{r}}{\xx{p}}}} &+&\mbox{$
 (-1)^{g_2 g_3}$ \nbr{\xx{r}}{\nbr{\xx{p}}{\xx{q}}} $=0,$}  \nonumber
\end{eqnarray}
where $g_1=p-1, \; g_2=q-1$ and $g_3=r-1$.  Therefore, the bracket
defined in eq.(19) may be identified with the Schouten-Nijenhuis (SN)
bracket of multivector fields \cite{SN} and the set of $LH$ multivector
fields equipped with the SN  bracket constitutes a {\bf Z}-graded Lie
algebra.\\


\newcommand{\inserta}{It should be noted that, unlike mechanics, in field
theory we have a nontrivial set of {\em primitive} Hamiltonian
multivector  fields \xxi{p}{0} satisfying
\[\xxi{p}{0} \inn\ \Omega = 0,       \]
$p=1,...,n.$ Evidently they constitute the subalgebra $\cal X$$_{0}$ of
the algebra $\cal X$ of Hamiltonian multivector fields w.r.t. SN
brackets and the map in eq.(18) actually maps the Hamiltonian forms
\ff{q} to the {\em equivalence classes} of Hamiltonian multivector
fields [\xx{p}] with respect to the addition of the primitive
Hamiltonian p-vector fields: [\xx{p}]=[\xx{p}+\xx{p}$_{0}$].
Therefore, the quotient algebra $\cal X / \cal X$$_{0}$ is more
adequate field theoretical analogue  of the Lie algebra of Hamiltonian
vector fields in mechanics than the original algebra $\cal X$ (cf. also
the related discussion in \cite{Kij ea}). \\   }


Now, taking \xxi{p}{1}\ and \xxi{q}{2} to be Hamiltonian fields, one
gets from eqs. (18) and (19):
\begin{eqnarray}
\mbox{\nbrpq{p}{q}\inn\ $\Omega$} & =\, &
                 \mbox{\lieni{p}{1}$d^{v}$\ff{s}$_{2}$} \nonumber \\ &=\, &
                 \mbox{$ (-1)^{p+1}\, d^{v}($\xxi{p}{1}\inn\
$d^{v}$\ff{s}$_{2})$} \\ & =:  &  \mbox{$-
d^{v}$\pbr{\ff{r}$_{1}$}{\ff{s}$_{_2}$},} \nonumber
\end{eqnarray}
where $r=n-p$ and $s=n-q$. The last equality in eq.(21) defines the
analogue of the {\em Poisson bracket} acting on Hamiltonian forms of
various degrees. As it is seen from the definition, it is related to
the Schouten-Nijenhuis bracket of  multivector Hamiltonian fields and
the polysymplectic $(n+1)$-form \oo\ in just  the same way as the usual
Poisson bracket is related to the Lie bracket of Hamiltonian vector
fields and the symplectic form. In eq.(21) we have actually shown that
the S-N bracket of two Hamiltonian multivector fields is Hamiltonian
(see the second equality), and then postulated in the third equality
that the Hamiltonian form corresponding to the S-N bracket of two
Hamiltonian fields defines the Poisson bracket of the Hamitonian forms
they are associated with according to the map given by eq.(18). \\

    The degree counting in eq. (21) gives
\beq
\mbox{$deg$\pbr{\ff{r}$_1$}{\ff{s}$_2$} $= r+s-n+1$}
\eeq
and with the help of eq.(20) one also finds
\beq
\mbox{\pbr{\ff{r}$_1$}{\ff{s}$_2$} $= -(-1)^{\sigma}$
\pbr{\ff{s}$_2$}{\ff{r}$_1$}, }
\eeq
where $\sigma=(n-r-1)(n-s-1)$.  From the definition in eq. (21) the
following useful formulae for the Poisson bracket can also be obtained:

\beq
\mbox{\pbr{\ff{r}$_1$}{\ff{s}$_2$} = $(-1)^{(n-r)}X_{1} \inn\ d^{v}$
\ff{s}$_2$ = $(-1)^{(n-r)}$$X_{1} \inn\ X_{2} \inn\  \Omega.$    }
\eeq \\
These equations resemble  familiar definitions of the Poisson bracket
in mechanics, but they are merely a consequences, as in mechanics, of
the fundamental definition, eq. (21), which is directly related to the
basic symmetry principle, eq. (11). Note, that in spite of the fact
that the definition in eq. (21) determines only a vertical exterior
derivative  of the Poisson bracket  there is no arbitrariness "modulo
exact form" in the definition of the Poisson bracket itself (cf.
\cite{Gold+Stern,Herm lie,Kij ea})  since the latter is required to map
the horizontal forms to horizontal ones  while the  $d^v$-exact
addition would necessarily be vertical. \\

\inserta

By a straighforward calculation one can obtain the following properties
of the Poisson brackets: (i) the graded analogue of the Leibniz rule
\beq
\mbox{\pbr{\ff{p}}{\ff{q} $\wedge$ \ff{r}} $=$ \pbr{\ff{p}}{\ff{q}} $\wedge$
\ff{r} $+ (-1)^{q(n-p-1)}$ \ff{q} $\wedge$ \pbr{\ff{p}}{\ff{r}}      }
\eeq
which means that the Poisson bracket with a $p$-form acts as a {\em
graded derivation} of degree $(n-p-1)$ (see also eq. (22)), and also
(ii) the graded Jacobi identities:
\begin{eqnarray}
\mbox{$(-1)^{g_1 g_3}$\pbr{\ff{p}}{\pbr{\ff{q}}{\ff{r}}}} &+& \nonumber \\
\mbox{$(-1)^{g_1 g_2}$\pbr{\ff{q}}{\pbr{\ff{r}}{\ff{p}}}} &+&
\mbox{$(-1)^{g_2 g_3}$\pbr{\ff{r}}{\pbr{\ff{p}}{\ff{q}}}$= 0,$}  \nonumber
\end{eqnarray}
where $g_1 = n-p-1$, $g_2 = n-q-1$ and $g_3 = n-r-1$.  Thus, the space
of Hamiltonian forms equipped with the Poisson bracket operation as
defined above constitutes a {\bf Z}-graded Lie algebra. Moreover, since
the exterior algebra of forms (i.e. the algebra w.r.t. the
$\wedge$-product) is itself a {\bf Z}-graded supercommutative
associative algebra, one can conclude that the Hamiltonian forms
constitute the so-called {\em Gerstenhaber algebra} \cite{Gerstenhaber}
(see also
\cite{LiZuck93}) with respect to our Poisson brackets and the exterior
product of forms.

\section{Equations of motion of Hamiltonian $(n-1)$-forms and the DW canonical
equations in the bracket form.}

In this section we consider the operation on the Hamiltonian forms
which is generated by the  Poisson bracket with the DW Hamiltonian
function and show how the equations of motion of $(n-1)$-forms can be
written with the help of this bracket. Then we shall shortly discuss
the field theoretic analogues of the integrals of motion and the
suitable choise of the canonically conjugate variables.  \\

{}From the degree counting in eq. (22) we see that only the $(n-1)$-forms
have nonvanishing brackets with $H$ and these are the 0-forms. Let us
calculate the bracket of the general Hamiltonian (n-1)-form
\[F:= F^{i}\der_{i}\inn\ \widetilde{vol}\]
with the DW Hamiltonian function $H$:
\beq
\mbox{\pbr{$F$}{$H$} $= -$ \xxi{1}{F}\inn\ $d^{v}H.$  }
\eeq
The components of the vector field \xxi{1}{F}$:= X^a\der_a +
X^i_a\der^a_i$ associated with $F$ are  to be calculated from the
equation
\beq
\mbox{\xxi{1}{F} \inn\ $\Omega= d^{v}${\em F} }  
\eeq
which reads in components
\beq
\mbox{$(-X^a dp^i_a + X^i_a dy^a)\wedge \dvol{i}
= (\der_a F^i dy^a + \der ^a_j F^i dp^j_a)\wedge \dvol{i}$ } \nonumber
\eeq
and yields

\beq
X^{a} \delta^{i}_{j}=-\der^{a}_{j}F^{i},
\eeq
\beq
X^{i}_{a}=\der_{a}F^{i}.
\eeq

 Hence, in contrast with mechanics, no arbitrary $(n-1)$-forms can be
Hamiltonian (i.e. to ensure the consistency of both sides of eq. (27)
and to give rise to some Hamiltonian vector field), but only those
which satisfy the condition (29) which restricts the dependence of the
components of $F$  on the DW canonical momenta $p^i_a$.  For such
$(n-1)$-forms  one has:

\beq
\pbr{H}{F} = \der_a F^i \der^a_i H + X^a \der_a H.
\eeq
Now,  the {\em total}\ (i.e. taken on extremals) exterior differential
{\bm $d$} of {\em F}
\[\mbox{{\bm $d$}}F:= (\der_a F^j \der_i y^a + \der^a_k F^j \der_i p^k_a
+ \der_i F^j ) dx^i \wedge \dvol{j} \]
on account of the condition (29) takes the form
\[\mbox{{\bm $d$}}F= (\der_a F^i \der_i y^a - X^a \der_i p^i_a
+ \der_i F^i) \vol. \]
Thus, with  the help of the DW Hamiltonian field equations, eqs. (7,8),
for an arbitrary Hamiltonian $(n-1)$-form $F$ one obtains:
\beq
\mbox{{\bm $d$}$F = $ \pbr{$H$}{$F$} $\widetilde{vol} + d^{hor} F$. }
\eeq
The last term $d^{hor}${\em F}$ = (\der_{i}F^{i})\widetilde{vol}$ in
eq.(32) appears for forms having explicit dependence on the space-time
variables.  The inverse Hodge dual of eq.(32)\footnote{Recall that
$\star \vol = \sigma$, where $\sigma=+1$ for Euclidean and $\sigma=-1$
for Minkowski signature of the metric; $\star^{-1}\, \star := 1$,
therefore on $n$-forms $\star^{-1} =
\sigma \star$ or, in general, $\star^{-1} = \sigma (-1)^{p(n-p)} \star$
on $p$-forms.}
\beq
\mbox{$\star$$^{-1}${\bm $d$}$F =$ \pbr{$H$}{$F$} + $\der_{i}F^{i}$}
\eeq
shows that the Poisson bracket with the DW Hamiltonian function is
related to the Hodge dual of the exterior differental in  essentially
the same way as  the time derivative is related to the Poisson bracket
with Hamilton's canonical function in mechanics.\\

Equation (33) contains, as a special case, the entire set of the DW
Hamiltonian field equations, eqs.(7,8). For, on account of $d^{hor} y^a
= 0$ and $d^{hor} p^i_a = 0$, by substituting $p_a := p^i_a \dvol{i}$
and then $y^a_i := y^a \dvol{i}$ for the $(n-1)$-form $F$, from eqs.
(30) and (32) we obtain

\[\mbox{$\star^{-1}${\bm $d$}} p_a = \pbr{H}{p_a} = - \der_a H \]

and

\[\mbox{$\star^{-1}${\bm $d$}} y^a_i = \pbr{H}{y^a_i} = \der^a_i H. \]

  Note that the dual of the total exterior derivative
$\star^{-1}$\mbox{{\bm $d$}} in eq. (33) is in fact nothing else than
the generalized Lie derivative with respect to the total n-vector field
$\xx{n}$$^{tot}$ = $\xx{n}$$^{v}$ + $\xx{n}$$^{hor}$ annihilating
$\Omega_{DW}$ (see eqs.(4)-(6) above).  The term including the Poisson
bracket is due to the Lie derivative w.r.t. the vertical part
$\xx{n}$$^{v}$ of this n-vector field while the last term in eq. (28)
is due to the Lie derivative w.r.t. its horizontal part
$\xx{n}$$^{hor}$; thus, $\star^{-1}${\bm $d$} = $(-1)^n{\cal
L}$$_{\stackrel{n}{X}^{_{_{tot}}}}$ when acting on the Hamiltonian
(n-1)-forms. In the subsequent section we will discuss how  this
bracket representation of the equations of motion may be generalized to
the Hamiltonian forms of degrees $p\leq(n-1)$.\\

Equations of motion in the bracket form suggest a natural
generalization of the classical notion of an integral of motion to
field theory.  Let $\cal J$ be the Hamiltonian $(n-1)$-form which does
not depend explicitly on space-time coordinates and has vanishing
Poisson bracket with the DW Hamiltonian function. Then from eq. (32)
the {\em conservation law} follows:
\[\mbox{{\bm $d$} $\cal J$$=0$.}\]
Thus, the field theoretical analogues of integrals of motion in
the present formulation are the $(n-1)$-forms corresponding to
conserved currents. Like the conserved quantities in mechanics they are
characterized by the condition
\beq
\mbox{\pbr{$\cal J$}{$H$} $= 0$.}
\eeq
Taking $\cal J$$_1$ and $\cal J$$_2$ to be the $(n-1)$-forms satisfying
eq.(34) and using the Jacobi identities, eq.(27), one gets
\[\mbox{\pbr{$H$}{\pbr{$\cal J$$_1$}{$\cal J$$_2$}}$\,=0$}.\]
Therefore, the Poisson bracket of two consereved currents fulfilling
eq.(34) is again a conserved current of the same kind. Latter statement
extends to field theory the {\em Poisson theorem} \cite{Arnold} that
the Poisson bracket of two integrals of motion of a Hamiltonian flow is
again an integral of motion. One can also conclude that the set of
conserved $(n-1)$-form currents having vanishing Poisson  bracket with
the DW Hamiltonian function is closed with respect to the Poisson
bracket and thus forms a Lie algebra being a subalgebra of the graded
algebra of all Hamiltonian forms.

    Furthermore, eq.(34) means that the Lie derivative of $H$ w.r.t. a
vertical  vector field $X$ associated with $\cal J$ vanishes, i.e. $H$
is invariant w.r.t. a symmetry generated by $X$ and $\cal J$ is a
conserved current corresponding to this symmetry of the DW Hamiltonian.
We  arrive thus at a sort of a field theoretical extension of the {\em
Hamiltonian Noether theorem} (cf. for example \cite[a]{Arnold} $\S$40
or \cite[b]{Arnold} $\S$15.1). Note that this extension concerns only
the symmetries  generated by the {\em vertical} vector fields.    \\

   The way in which the canonical Hamiltonian field equations are
represented in terms of the Poisson brackets sheds light on the
question as to which variables may be considered in the present
formalism as the canonically conjugate ones. As we know from mechanics,
canonically conjugate variables (i) have  "simple" mutual Poisson
brackets (leading to the Heisenberg algebra structure) and (ii) their
products have the dimension of action.  It is easy to see that in our
approach the pair of variables
\beq
(y^a, p_a := p^i_a \dvol{i}),
\eeq
one of which is $0$-form and another is $(n-1)$-form, may be considered
as a pair of canonically conjugate variables. The Poisson brackets of
these variables
\beq
\pbr{y^a}{p_b} = - \delta^a_b, \; \pbr{y^a}{y^b}=0, \; \pbr{p_a}{p_b}=0
\eeq
turn out to be the same as those of coordinates and canonically
conjugate momenta in mechanics. Indeed, from eqs. (23),(24) one obtains
\[\pbr{y^a}{p_b}=-\pbr{p_b}{y^a}=\mbox{\xx{1}$_{(p_b)}$} \inn\, dy^a =
f-\delta^a_b, \] where one has used in the last equality that the
vector field \xx{1}$_{(p_b)}$ associated with the $(n-1)$-form $p_b$ is
given by
\[\mbox{\xx{1}$_{(p_b)}$} \inn\, \Omega = dp_b , \]
so that
\[\mbox{\xx{1}$_{(p_b)}$}=-\der_b  \]
(cf. also eqs. (29),(30)).\\

It should be noted, however, that in principle this choice is not
unique.  For example, we could also choose the pair $(y^a \dvol{i},
p^j_b)$: the (nonvanishing) Poisson bracket in this case is also
remarcably simple, namely         
\beq
\pbr{y^a \dvol{i}}{p^j_b} = - \delta^j_i\, \delta^a_b .
\eeq
Such a freedom is due to the "canonical supersymmetry", eq.(17), mixing
the forms of different degrees. It might be especially useful in field
theories in which the field variables themselves are forms, like a
1-form potential $A_{\nu}dx^{\nu}$ in electrodynamics or a  2-form
potential in the Kalb-Ramond field theory: to the $p$-form field
variable the $(n-p-1)$-form conjugated momentum may be associated, and
their mutual Poisson bracket may be shown (cf. Sect. 6.2) to be equal
to one (up to a sign). Note that the extension of our construction in
the following section, which involves the $n$-forms as Hamiltonian
forms, provides in principle still more freedom in specifying the
canonically conjugate variables.   \\

\section{Equations of motion of Hamiltonian forms of an arbitrary degree}

   In Sect. 3 we have found that  the proper field theoretical
generalization of the Hamiltonian functions of mechanics are the
horizontal forms of various degrees from $0$ to $(n-1)$, on which the
analogue of the Poisson bracket operation involving the forms of
various degrees was defined. However, in Sect. 4 only the equations of
motion of $(n-1)$-forms were formulated in terms of these generalised
Poisson brackets.  It seems natural to ask whether this circumstance is
due to some priveleged position of $(n-1)$-forms in the formalism
(which might indeed be the case since some of them yield classical
observables after integrating over the spacelike surface orthogonal to
the time-direction of an observer) or there exists a possibility to
write in bracket form the equations of motion of Hamiltonian forms of
any degree. In this Section we argue that the second alternative may be
realized indeed by means of a slight generalization of the construction
of Sect. 3 leading to further extension of the notion of Hamiltonian
forms and the associated Hamiltonian fields.  \\

The problem we meet trying to extend the equation of motion of
$(n-1)$-forms to Hamiltonian forms of arbitrary degree $p < (n-1)$ is
essentially that $\pbr{\ff{p}}{H}$ vanishes {\sl identically} when
$p<(n-1)$. The possible way out of that is suggested by the observation
that for all $p$ the bracket $\pbr{\ff{p}}{H\vol}$ would not vanish, as
the formal degree counting based on eq. (20) indicates, if one could
extend our hierarchy of equations relating the Hamiltonian forms and
the Hamiltonian multivector fields, eq. (18), so as to supplement the
set of Hamiltonian forms with the horizontal forms of degree $n$, as
the form $H\vol$ is. This is possible indeed, if the object
$\tilde{X}^{v}$ which one associates with the horizontal n-form
$\ff{n}$ by means of the map
\beq
\mbox{$\tilde{X}$$^{v}_{F}$ \inn\ $\Omega^{v} = d^{v}$ \ff{n} }
\eeq
is thought to be the vertical-vector-valued horizontal one-form
\beq
\tilde{X}^{v} := X^{V}_{\cdot\, k}\, dx^{k} {\bf \otimes}\, \der_{V},
\eeq
and the
inner product \inn\ to be the Fr\"{o}licher-Nijenhuis (FN) inner
product of a vector-valued form and a form \cite{FN,Nono}:
\beq
\mbox{$\tilde{X}$\inn\ \oo\ := $X^{V}_{\cdot\, k} dx^{k}$ $\wedge\, (\der_{V}$
\inn\ \oo).  }
\eeq
Here we use the usual symbol of the inner product of vectors and forms
and imply that the tilde over the argument at the l.h.s. indicates that
it is a vector-valued form so that \inn\ in this case denotes the  FN
inner product of a vector-valued form and a form. \\


By extending formulae (24) one can define now the {\em left} Poisson
bracket of the  p-form \ff{p}  with  the n-form \ff{n} as follows:
\beq
\mbox{\pbr{\ff{n}$_1$}{\ff{p}$_2$}
= $\tilde{X}^{v}_{F_1}$ \inn\ $d^v$ \ff{p}$_2$. }
\eeq
This expression may be substantiated by the considerations similar to
those which led from eq. (17) to eq. (24), provided one supplements the
hierarchy of symmetries in eq. (17) with the additional assumption
\beq
\mbox{$\cal L$$_{\tilde{X}^{v}}$$\Omega = 0$ }
\eeq
formally corresponding to $p=0$
and defines the generalized Lie derivative of an arbitrary form $\omega$
with respect to the vertical-vector-valued  form $\tilde{X}^{v}$  as
\beq
\mbox{$\cal L$$_{\tilde{X}^{v}}$$\omega := \tilde{X}^{v} \inn\ d^{v}\omega -
d^{v}(\tilde{X}^{v} \inn\ \omega).$  }
\eeq
Note that $\cal L$$_{\tilde{X}^{v}}$ maps  p-forms to (p+1)-forms.   \\

 Taking $\ff{n} = H\vol$, the components  of the associated
vector-valued form $\tilde{X}$$_H$ may be found from eq. (38) to be
\beq
\mbox{$\tilde{X}$$^a_{\cdot k}$$ = \der^a_k H,$ \hspace*{1em}
$\tilde{X}$$^i_{a k}$$\delta^k_i = -\der_a H. $}
\eeq
We see from eq.(44) that $\tilde{X}$$_H$ is also suitable generalization,
as $\xx{n}$ is, of the canonical Hamiltonian vector field in mechanics
because using the natural parametrization of $\tilde{X}$$_H$
\beq
\tilde{X}^V_{\cdot k} = \frac{\der z^V}{\der x^k}
\eeq
leads to
the DW Hamiltonian field equations again. It is obvious that
the horizontal counterpart of  the vertical-vector-valued form
$\tilde{X}^{v}$$_H$ associated with $H\vol$ is
\[ \mbox{$\tilde{X}$$^{hor}$$ = \delta^i_k\,dx^k \otimes \der_i \, .$} \]

Now, the total exterior differential of the $p$-form has the form

\[\mbox{{\bm $d$} \ff{p}$ = dx^k\wedge\der_k z^V \der_V$\ff{p}$ + d^{hor}$
\ff{p}   }  ,\]
and from eq. (41) it follows

\beq
\mbox{\pbr{$H\vol$}{\ff{p}}$=\tilde{X}^V_{\cdot k} dx^k \wedge \der_V$
\ff{p} .}
\eeq
Thus the DW canonical equations encoded in eqs. (44),(45) imply the
following equation of motion of an arbitrary $p$-form

\beq
\mbox{{\bm $d$}\ff{p}=\pbr{$H\vol$}{\ff{p}}$+d^{hor}$\ff{p} .}
\eeq
The l.h.s. of eq. (47) generalizes the total time derivative in the
equations of motion of a dynamical variable in mechanics, whereas its
last term generalizes the partial time derivative. It is clear from the
definition, eq. (43), that {\bm $d$}=$\cal L$$_{\tilde{X}^{tot}}$,
where $\tilde{X}^{tot}=\tilde{X}^v_H + \tilde{X}^{hor}$. \\

It should be noted that enlargening the set of the Hamiltonian
multivector fields of Sect. 3 by the vector-valued one-forms associated
with the Hamiltonian $n$-forms implies certain extension of the algebra
of Hamiltonian (and $LH$) fields w.r.t. the SN bracket. Let us closer
look at this extension. First, one defines the bracket of the
vector-valued-one-forms and  multivectors:     
\beq
\mbox{\nbr{$\tilde{X}$}{\xx{p}} \inn\ $\Omega:=$$\cal L$$_{\tilde{X}}$
(\xx{p} \inn\ $\Omega)$   .  }
\eeq
Therefore,
\beq
\mbox{\nbr{$\tilde{X}$}{\xx{p}} $\in \Lambda$$^p_1$,}
\eeq
where $\Lambda^p_q$ denotes the space of vertical-$p$-vector-valued
horizontal $q$-forms. Second, the bracket of the vector-valued forms
would be natural to define by the equality
\beq
\mbox{\nbr{$\tilde{X}$$_1$}{$\tilde{X}$$_2$} \inn\ $\Omega :=$
$\cal L$$_{\tilde{X}_{1}}$$(\tilde{X}_{2} \inn\ \Omega)$, }
\eeq
however, its r.h.s.  is identically zero, as it follows from the formal
degree counting. Hence, it does not actually define the bracket, but
only indicates that
\beq
\mbox{\nbr{$\tilde{X}$$_1$}{$\tilde{X}$$_2$} $\in \Lambda^1_2.$ }
\eeq
This is exactly the property which the Fr\"{o}licher-Nijenhuis bracket
of two vector-valued forms has \cite{FN,Nono}. Thus, it is natural to
identify the bracket of two vector-valued forms with the FN bracket. As
a result, the closure of the algebra will involve the vector-valued
$p$-forms of all degrees $p\leq n$ because
\beq
\mbox{\nbr{$\tilde{X}$$_1$}{$\tilde{X}$$_2$}$_{FN}$ $\in \Lambda^1_{p+q}$  }
\eeq
if $\tilde{X}$$_{1} \in \Lambda^1_p$ and $\tilde{X}$$_{2} \in
\Lambda^1_{q}$.  Notice, that an appearance the of vector-valued forms
of higher degrees does not lead to any extension of the algebra of
Hamiltonian forms: for $\tilde{X} \in \Lambda^1_{p>1}  $
\beq
\mbox{$\tilde{X}$ \inn\ $\Omega = 0$, }
\eeq
where $\inn\ $ denotes the FN inner product of a vector-valued q-form
$\tilde{X}$ and a form $\omega$, defined as follows:
\beq
\tilde{X}\inn\ \omega := X^{V}_{\cdot k_1 ... k_q} dx^{k_1}\wedge...\wedge
dx^{k_q} \wedge(\der_{V}\inn\ \omega).
\eeq
It follows from eq. (53) that  the vector-valued forms of the degree
higher than one extend only the subalgebra of {\em primitive}
Hamiltonian fields (see Sect.3).  \\

   Let us return now to the bracket in eq.(49): the bracket of a
vector-valued form and a $p$-vector gives a $p$-vector-valued form.
According to our scheme one should associate these objects with the
"Hamiltonian" (here in vague sense) forms via the polysymplectic form,
and then to define somehow the corresponding brackets. However, the
first task meets the problem of unique definition of the inner product
of $\tilde{X} \in \Lambda^p_1$ with forms while the second one leads to
the related problem of an appopriate definition of the "Lie derivative"
w.r.t. these $\tilde{X}$-s.  Moreover, since "most probably" the mutual
brackets (yet to be properly defined) of those $\tilde{X}$-s will yield
the elements from all the spaces $\Lambda^p_q$, we actually have to
solve the same problems for arbitrary vertical-multivector-valued
forms.  Thus, the problem is essentially to construct a graded algebra
of multivector-valued forms equipped with some appropriate bracket
operation generalizing both the Schouten-Nijenhuis and the
Fr\"{o}licher-Nijenhuis bracket. This is in fact the "well-known"
mathematical problem. However,  recently A.M.  Vinogradov has published
his "unification theorem" \cite{Vinogr} which states that SN and FN
algebras may be imbedded in certain {\bf Z}-graded quotient algebra of
the algebra of super-differential operators on the exterior algebra of
forms.  Although this result sounds highly relevant, the solution of
the problem outlined above, which would be satisfactory for our
purpose, as yet is not obtained by the author.

\section{Several simple applications}

\subsection{Interacting scalar fields}

As a simplest example of how the formalism we have constructed works,
let us consider the system of interacting real scalar fields
$\{\phi^{a}\}$ described by the Lagrangian density
\beq
L= \frac{1}{2}\der_i \phi^a \der^i \phi_a - V(\phi^a).
\eeq
The DW canonical momenta are

\beq
p^i_a := \frac{\der L}{\der(\der_i \phi^a)} = \der^i \phi_a
\eeq
and for the DW Hamiltonian function we easily obtain
\beq
H= \frac{1}{2}p^i_a p^a_i + V(\phi).
\eeq
In terms of the canonically conjugate (in the sense of Sect. 4)
variables $\phi^a$ and $\pi_a := p^i_a \dvol{i}$ which have the
following nonvanishing mutual Poisson bracket
\beq
\mbox{\pbr{$\phi^a$}{$\pi_b$} $ =-\delta^a_b$, }
\eeq
we can also write
\beq
H\vol = \frac{1}{2}\sigma(\star \pi^a) \wedge \pi_a + V(\phi)\vol.
\eeq
Finally, the canonical DW equations may be written in the bracket form

\begin{eqnarray}
\mbox{{\boldmath $d$}}\pi_a  &= \pbr{H\vol}{\pi_a}&=-\der_a H\vol, \nonumber \\
                             &                    &               \\
\mbox{{\boldmath $d$}}\phi^a &= \pbr{H\vol}{y^a}  &=\sigma\star
  \pi^a , \nonumber
\end{eqnarray}
which is equivalent to the field equations following from the Lagrangian (55):
\beq
\Box \phi_a = -\der_a V.
\eeq

\subsection{The electromagnetic field}

Let us start from the conventional Lagrangian density
\beq
L = -\frac{1}{4}F_{ij}F^{ij} - j_{i}A^{i},
\eeq
where $F_{ij}:=\der_i A_j - \der_j A_{i}$. For the canonical DW momenta
we get
\beq
\pi^{i}_{\cdot j}:= \frac{\der L}{\der(\der_i A^j)}= -F^{i}_{\cdot j}
\eeq
whence the primary constraints
\beq
\pi^{ij}+\pi^{ji}=0
\eeq
follow. Despite the DW Legendre transformation is singular, we can
define the canonical DW Hamiltonian function as usual:
\beq
H= -\frac{1}{4}\pi_{ij} \pi^{ij} - j_{i}A^{i}.
\eeq
However, due to the constraints using this Hamiltonian in the DW
Hamiltonian field equations leads to the incorrect equation
$\der_{i}A^{j}=\der H / \der \pi^{i}_{\cdot j} = \pi_{i}^{\cdot j}$:
only its antisymmetric part  is right. Usually the problems of this
sort are handled by substituting the constraints with some Lagrange
multiplies to the canonical Hamiltonian function and then applying the
well known Dirac's procedure. Our formalism offers another possibility
based on a freedom in choosing the canonically conjugate variables
which we have already mentioned in Sect.4.

    Namely, let us try to use as a canonical field variable the
one-form $\alpha$$=A_{i}dx^{i}$ instead of the set of its components
$\{A_{i} \}$. Then the canonically conjugate momentum may be found to
be the $(n-2)$-form
\[\pi := -F^{ij}\der_i \inn \der_j \inn \vol=\pi^{ij}\der_i \inn
\der_j \inn \vol.\]
To see this let us calculate the Poisson bracket of the 1-form $\alpha$
and the $(n-2)$-form $\pi$. Remark first, that the form $\pi$ is
Hamiltonian form as opposite to its dual 2-form $F_{ij}dx^i \wedge
dx^j$ which we might naively try to associate with $\alpha$  as its
conjugate momentum; moreover, the bracket of the latter two forms would
vanish for $n>4$, as a simple degree counting shows (see eq.  (22)).
Further, the components of the $(n-1)$-vector field $X_{\alpha}$
associated with $\alpha$  are defined by
\[X_{\alpha} \inn \Omega = d\alpha = dA_i \wedge dx^i,  \]
where
\[\Omega = -dA^i \wedge d\pi_i^{ j} \wedge (\dvol{j}).  \]
For the only nonvanishing component of $X_{\alpha}$ we get
\beq
X_i^{ ji_1 ... i_{n-2}} \varepsilon_{i_1 ... i_{n-2}jk} = g_{ik},
\eeq
where the first column of indices is a single index corresponding to
the direction $\der^i_j := \frac{\der}{\der \pi_{\cdot i}^{j}}$ in the
tangent space of the DW phase space.  The Poisson bracket of $\alpha$
and  $\pi$ is easily obtained from its definition
\beq
\mbox{\pbr{$\alpha$}{$\pi$} $ = (-1)^{n-1} X_{\alpha} \inn d\pi = -1$. }
\eeq
This property justifies our choice of the canonically conjugate
momentum of the one-form potential $\alpha$.  \\

In terms of new canonical variables $\alpha$ and $\pi$ the DW
Hamiltonian $n$-form is expressed as
\beq
H\vol = -\frac{1}{4}\sigma \pi \wedge (\star \pi) - \alpha \wedge j,
\eeq
where $j:= j^i \dvol{i}$  is the current density $(n-1)$-form.
Now, the Maxwell equations acquire the following Hamiltonian form in
terms of new variables and the Poisson brackets:
\begin{eqnarray}
\mbox{{\boldmath $d$}$\alpha \;=\; $\pbr{$H\vol$}{$\alpha$}}&=&\star^{-1}\pi,
  \nonumber \\
                                                            & &      \\
\mbox{{\boldmath $d$}$\pi \;= \; $\pbr{$H\vol$}{$\pi$}}     &=&j. \nonumber
\end{eqnarray}
Thus we have obtained a covariant Hamiltonian formulation of Maxwell's
electrodynamics without recourse to the formalism of the fields with
constraints. The constraints, both gauge and initial data, which, of
course, did not disappear nowhere  can be taken into account {\em
after} the covariant Hamiltonian formulation was constructed.

\subsection{The Nambu-Goto string}

\newcommand{\xdot}{\stackrel{.}{x}}
\newcommand{\xprim}{x'}

The classical dynamics of a string sweeping in space-time the world-sheet
$x^a = x^a (\sigma,\tau)$ is determined by the Nambu-Goto Lagrangian
\beq    
L= -T\sqrt{(\xdot \cdot \xprim)^2 - \xprim^2 \mbox{$\xdot$$^2$}} =
-T\sqrt{-det\|\der_i x^a \der_j x_a \|},
\eeq
where $\xdot^a = \der_{\tau} x^a, \xprim^a = \der_{\sigma} x^a$, $T$ is a
string rest tension and we have also used the following notation for the
world-sheet parameters $(\sigma,\tau) = (\tau^0 , \tau^1):=(\tau^i); \; i=0,1$.

Define the canonical DW momenta:
\begin{eqnarray}
p^0_a := \frac{\der L}{\der \mbox{$\xdot$$^a$}}  = T^2\frac{(\xprim
 \cdot \xdot)\xprim_a - \xprim^2 \xdot_a}{L},& & \nonumber \\
      &                                 &           \\
p^1_a := \frac{\der L}{\der \xprim^a} = T^2\frac{(\xprim
 \cdot \xdot)\xdot_a - \mbox{$\xdot$$^2$} \xprim_a}{L}.& &   \nonumber
\end{eqnarray}
{}From eqs.(71) the following identities follow
\begin{eqnarray}
p^0_a \xprim^{a}=&0, \hspace*{4em} (p^0)^2 + T^2 \xprim^2&= 0, \nonumber \\
                 &                                       &    \\
p^1_a \xdot^{a} =&0, \hspace*{4em} \mbox{$(p^1)^2 + T^2 \xdot$$^2$}&= 0,
 \nonumber
\end{eqnarray}
however, they do  {\em not} have a meaning of the Hamiltonian
constraints within the DW canonical formalism since they do not imply
any relations between the generalized coordinates $x^a$ and the
generalized momenta $p^i_a$. In fact, with the help of these identities
the eqs. (71) may be easily solved (if $L \neq 0$) yielding the
expression of the generalized velocities $(\xdot, \xprim)$ in terms of
the DW momenta; it proves {\em de facto} that the DW Legendre transform
for the Nambu-Goto Lagrangian is regular!  In terms of the DW momenta
the DW Hamiltonian function takes the form
\beq
H=-\frac{1}{T}\sqrt{-det\|p^i_a p^{aj}\|}
\eeq
and can also be expressed in terms of the 1-form momentum variables
\[\pi_a := p^i_a \varepsilon_{ij}d\tau^j \]
canonically conjugate (in the sense of Sect. 4) to $x^a$.
It is easily checked that
\[det\|p^i_a p^{aj}\|=\frac{1}{2}(\varepsilon_{ij}p^i_a p^j_b)
(\varepsilon_{ij}p^{ai}p^{bj})  \]
and
\[\varepsilon_{ij}p^i_a p^j_b =  \star(\pi_a \wedge \pi_b).  \]
The string equations of motion in terms of the Poisson brackets can be written
now as
\begin{eqnarray}
\mbox{{\boldmath $d$}$x^a  $}&= \mbox{ \pbr{$H\vol$}{$x^a$}}&=
             \;\frac{\der H}{\der p^i_a}d\tau^i, \nonumber \\
             & &   \\
\mbox{{\boldmath $d$}$\pi_a$}&= \mbox{ \pbr{$H\vol$}{$\pi_a$}}&= \;0. \nonumber
\end{eqnarray}

As yet another application we show how the Poincar\'{e} algebra is
reproduced with the help of our brackets. In the $x^a$-space the
translations are generated by the vector fields $X_a := \der_a$ and the
Lorentz rotations by the bivectors $X_{ab} := x_a \der_b - x_b \der_a$.
The corresponding conserved current densities are the one-forms:
\beq
\mbox{$\pi_a$$  \;\;$ {\rm and} $\;\; $$\, \mu_{ab}:=x_a \pi_b - x_b \pi_a,$ }
\eeq
and from the string equations of motion it follows
\beq
\mbox{\boldmath $d$} \pi_a = 0 \;\;  \mbox{\rm and} \;\; \mbox{\boldmath $d$}
\mu_{ab} = 0.
\eeq
Now, a straightforward calculation of the Poisson brackets of these
$1$-forms yields:
\begin{eqnarray}
\hspace*{2em}\pbr{\pi_a}{\pi_b}&=&0, \nonumber  \\
\pbr{\mu_{ab}}{\pi_c}          &=&g_{ac}\pi_b - g_{bc}\pi_a,   \\
\pbr{\mu_{ab}}{\mu_{cd}}       &=&C_{abcd}^{ef} \mu_{ef} \nonumber,
\end{eqnarray}
where $g_{ab}$ is the $x$-space metric and
\[C_{abcd}^{ef}=-g^e_cg^f_ag_{bd} + g^e_cg^f_bg_{ad} -
g^e_ag^f_dg_{bc}  + g^e_bg^f_dg_{ac} \] are the Lorentz group structure
constants. Thus the internal Poincar\'{e} symmetry of a string is
represented with the help of the Poisson brackets of 1-forms
corresponding to the conserved currents related to this symmetry.

\section{Discussion}

In this paper we have discussed a possible extension to the
finite-dimensional De Donder-Weyl Hamiltonian formulation of field
theory of some  of the structures of the classical Hamiltonian
mechanics, in particular, those which are known (or commonly believed)
to be important for canonical quantization.  Although the symplectic
form is explicitly needed only in the geometric quantization schemes,
we started first from its appropriate field theoretical (within the DW
framework) analogue since this  seems to provide the only reliable
basis for generalizations.  Unlike the symplectic $2$-form in mechanics
$(n=1)$ its field theoretical (in $n$ dimensions) generalization in the
present formalism, the polysymplectic $(n+1)$-form which is determined
from the Poincar\'{e}-Cartan canonical form, is not purely vertical. As
a result, it determines the map between  multivectors and forms of
various degrees, which generalizes the map between vectors and
functions given by the symplectic form in mechanics. However, our map
is rather a map between the {\em equivalence classes} of Hamiltonian
multivector fields modulo the addition of the {\em primitive}
Hamiltonian fields, which are
defined in Sect. 3, and the specific class
of forms, called the {\em Hamiltonian forms}; the latter term in
general implies certain limitation for the  dependence of forms on the
DW  canonical momenta.  The meaning of this limitation, which is
obtained only as an analytic relation following from the consistency of
the equation defining the components of the multivector associated with
a given form, is not quite clear.   \\

Introduction of the generalized Lie derivative of forms with respect to
the multivector fields enables us to define the Poisson brackets of
Hamiltonian forms which mix the forms of different degrees and are
shown to be connected with the Schouten-Nijenhuis brackets of
Hamiltonian multivector fields they are associated with. The brackets
can be both odd or even depending on the degrees of their arguments and
are proved to fulfil the graded derivation property and  the graded
Jacobi identity.  Thus, the set of Hamiltonian multivector fields and
the set of Hamiltonian forms possess a clear graded Lie algebra
structure and, moreover, the Gerstenhaber algebra structure. One of the
consequences of the "mixing property" of our brackets is that the
natural pairs of canonically conjugate field and momentum variables
consist typically of forms of different degrees whose components are
field variables or DW momenta. We also show that the Poisson bracket
with the DW Hamiltonian function generates the Hodge dual of the
exterior differential on the space of Hamiltonian $(n-1)$-forms,
whereas the bracket with the $n$-form $H\widetilde{vol} = \star H$
generates the exterior differential of an arbitrary (Hamiltonian) form.
This leads, in particular, to the representation of the De Donder-Weyl
Hamiltonian  field equations in the bracket form which is similar to
that in mechanics where the time derivative of a dynamical variable is
generated by the Poisson bracket with Hamilton's function.  \\

     However, the n-forms, as the Hamiltonian forms, are associated
with the vector-valued one-forms enlargening the space of Hamiltonian
multivector fields. As we have argued, this implies a certain extension
of the graded Lie algebra of Hamiltonian multivector fields.  The
algebraic closure of this extension involves the multivector-valued
forms of all possible degrees and requires  the appropriate definition
of some bracket operation for these objects, which would extend the
Lie, Schouten-Nijenhuis and Fr\"{o}licher-Nijenhuis brackets. This
problem is related to the problem of construction of the graded algebra
of multivector-valued forms covering both SN and FN graded algebras.
The techniques we have used in this paper did not allow us to construct
this superalgebra; perhaps the recent unification theorem by A.M.
Vinogradov  \cite{Vinogr} might be helpful in this connection. As a
speculation, one could expect that taking into consideration of all the
elements of this enlarged graded algebra can also lead to a certain
extension of the algebra of Hamiltonian forms and possibly will allow
us to weaken or avoid the restrictive analytical condition on the
Hamiltonian forms.  In other words, the question is whether one can
associate the objects of more general nature, the multivector-valued
forms, with the forms which are not Hamiltonian according to the
definition of Sect.3. An extension of the class of Hamiltonian forms
seems also to be desirable  in view of the fact that the (horizontal)
Hodge duals of the Hamiltonian forms which we seemingly need to
consider together with the Hamiltonian forms, as the examples in Sect.
6 indicate, are not Hamiltonian in general \cite{Kan-inprep}.  \\

   Let us sketch the connections of the formalism presented here with
the conventional instantaneous Hamiltonian formalism for fields (more
detailed treatment will be presented elsewhere). Let us choose a
space-like surface $\Sigma$ in the $x$-space (here we will assume it to
be pseudueuclidean with the signature $++...+-$). The restrictions of
the DW phase space variables  to $\Sigma$ will be the functions of
$x$-s. In particular, if $\Sigma$ is given by the equation $x^n = t$
($n$ is the number of the time-like component of
$\{x^i\}=\{x^1,...,x^{n-1},x^n\}$, not an index), we have
$y^a|_{\Sigma}=y^a({\bf x},t)$ and $p^i_a|_{\Sigma}=p^i_a({\bf x},t)$,
where {\bf x} denotes the space-like components of $\{x^i\}$. Moreover,
the restriction of forms to $\Sigma$ implies setting $dx^n=0$, so that
for $p_a:=p^i_a\der_i\inn \vol$ we have $p_a|_{\Sigma}=p^n_a({\bf
x},t)\der_n\inn \vol$, where $\der_n \inn \vol$ is obviously the
$(n-1)$-volume form on $\Sigma$, which we shall denote as $d\bf x$. The
functional symplectic 2-form $\omega$ on the phase space of the
instantaneous formalism may be related now to the restriction of the
polysymplectic form $\Omega^v$ to $\Sigma$ in the following way (cf.
\cite{Gotay multi2,Gotay ea}):
\[\omega=\int_{\Sigma}(\Omega^v|_{\Sigma})=
-\int_{\Sigma} dy^a({\bf x}) \wedge dp^n_a({\bf x}) d{\bf x}. \] Then,
the equal-time Poisson bracket  of $y^a({\bf x})$ with the canonical
conjugate momentum $p^n_a({\bf x})$:  
\[\{y^a({\bf x}),p^n_b({\bf y})\}_{PB}=\delta^a_b\delta({\bf x}-{\bf y}) \]
may be related to the Poisson bracket of the canonically conjugate
variables $y^a$ and $p_a$ of the DW theory (see Sect. 4) as follows:
\[\int_{\Sigma_x}\int_{\Sigma_y}\{y^a({\bf x}), p^n_b ({\bf y})\}_{PB}
f({\bf x})f({\bf y}) d{\bf x}d{\bf y} = -
\int_{\Sigma}\{y^a,p_b\} f({\bf x})d{\bf x},  \]
where $f({\bf x})$ is a test function. In general, one can anticipate
the following relationship between the generalized Poisson bracket of
Hamiltonian forms and the equal-time Poisson bracket of their
restrictions to the space-like surface $\Sigma$:
\[\int_{\Sigma_x}\int_{\Sigma_y} \phi_1 (x) \wedge
\{\mbox{\ff{p}$_{1}|_{\Sigma_x}(x),$\ff{q}$_2|_{\Sigma_y}(y) \}_{PB}
\wedge \phi_2 (y)$} \sim\, \int_{\Sigma} \phi_1 \wedge
\mbox{\pbr{\ff{p}$_1$}{\ff{q}$_2$}}\wedge \phi_2, \]
where $\phi_1$ and $\phi_2$ denote the "test forms" of degree $(n-p-1)$
and $(n-q-1)$ respectively and the standart Poisson bracket $\{\;
,\;\}_{PB}$ of forms is defined via the Poisson brackets of their
components.  This formula reproduces, in particular, the canonical
equal-time Poisson brackets from the generalized Poisson brackets of
the admissible pairs of canonically conjugate variables (see the end of
Sect. 4).  However, it fails to reproduce some of the Poisson brackets
of interest in field theory from the generalized Poisson brackets of
Hamiltonian forms, although all such examples known to the author are
related to the quantities which are not Hamiltonian forms. This is
another indication that our canonical scheme should be extended.    \\

It is interesting to note in conclusion that the algebraic structures
arised in our formalism are cognate with those appearing in the
BRST-inspired approaches in field theory, in particular, in the
antibracket formalism (see for example
\cite{Henneaux book}). The latter, of course, are established within the
functional framework which is conceptually different from the spirit of
this paper. Nevertheless, deeper relationship rather than a superficial
algebraic analogy may be expected in view of the connection discussed
above between the usual Poisson brackets and those suggested in this
paper; in this case one could hope to clarify the geometrical origin of
the BRST formalism. It is worthy of noting in this connection that the
Gerstenhaber algebra structure which we have found for graded Poisson
bracket algebra of Hamiltonian forms has appeared recently also in the
discussion of the BRST-algebraic structure of string theory
\cite{LiZuck93}.      \\

   {\bf Acknowledgments.} I am grateful to Professor H.A. Kastrup for
his kind reception at the Institute for Theoretical Physics, RWTH, Aachen,
discussions and stimulating criticism. My thinking about the topics of this
paper was initiated and has been stimulated by his Physics Reports review
\cite{Kastrup83}. I  thank Wulf B\"{o}ttger for his interest in this work,
stimulating discussions and useful comments. The financial support of the
above mentioned Institute is gratefully acknowledged.  I wish also thank
J. Marsden and M. Gotay for kindly making  (the August 1992 version of)
their book \cite{Gotay ea} available to me.

\renewcommand{\baselinestretch}{1.0}
\normalsize\large

\end{document}